\documentclass[3p,authoryear,sort&compress,times]{elsarticle}
\usepackage{amsmath,amsfonts}  
\usepackage{color}
\usepackage{lineno}
\usepackage{wasysym}
\usepackage{array}
\usepackage{longtable}
\usepackage{booktabs}
\usepackage{multirow}
\journal{Ecology Letters}

\usepackage{setspace}
\usepackage{setspace,caption}
\AtBeginCaption{\doublespacing}
\usepackage{ marvosym }

\usepackage{natbib}
\usepackage{etoolbox}
\usepackage{filecontents}
\newbool{MyRefNumbers}
\booltrue{MyRefNumbers} 

\usepackage{geometry}
\usepackage{pdflscape}

\begin{document}

\begin{center}
\LARGE{Edge fires drive the shape and stability of tropical forests}
\end{center}

\vspace{-.4cm}

\begin{center}
\large{Laurent H\'{e}bert-Dufresne$^{1,2}$, Adam F.\ A.\ Pellegrini$^{3,4}$ Uttam Bhat$^{2,5,6}$,\\  Sidney Redner$^{2}$, Stephen W.\ Pacala$^{4}$ and Andrew M.\ Berdahl$^{2,7\dagger}$}
\end{center}

\vspace{-.45cm}

\begin{enumerate}
{\setlength\itemindent{-.5cm}{\setlength\itemsep{-2pt}
\small{
\item Department of Computer Science and Vermont Complex Systems Center, University of Vermont, Burlington, VT 05405, USA
\item Santa Fe Institute, 1399 Hyde Park Road, Santa Fe, NM 87501, USA
\item Department of Earth System Science, Stanford University, Stanford, CA 94305, USA
\item Department of Ecology and Evolutionary Biology, Princeton University, Princeton, NJ 08544, USA
\item Department of Physics, Boston University, Boston, MA 02215, USA
\item School of Natural Sciences, University of California, Merced, Merced, CA 95343, USA
\item School of Aquatic and Fishery Sciences, University of Washington, Seattle, WA 98195, USA
}}}
\end{enumerate}

\vspace{-.3cm}


\vspace{-.3cm}

\noindent\makebox[\linewidth]{\rule{\textwidth}{0.4pt}}

\vspace{-.15cm}

\noindent {\bf{Abstract}}

\noindent

In tropical regions, fires propagate readily in grasslands but typically
consume only edges of forest patches. Thus forest patches grow due to tree
propagation and shrink by fires in surrounding grasslands. The interplay
between these competing edge effects is unknown, but critical in determining
the shape and stability of individual forest patches, as well the
landscape-level spatial distribution and stability of forests. We analyze
high-resolution remote-sensing data from protected areas of the Brazilian
Cerrado and find that forest shapes obey a robust perimeter-area scaling
relation across climatic zones. We explain this scaling by introducing a
heterogeneous fire propagation model of tropical forest-grassland
ecotones. Deviations from this perimeter-area relation determine the
stability of individual forest patches. At a larger scale, our model predicts
that the relative rates of tree growth due to propagative expansion and
long-distance seed dispersal determine whether collapse of regional-scale
tree cover is continuous or discontinuous as fire frequency changes.

\vspace{-.1cm}

\noindent\makebox[\linewidth]{\rule{\textwidth}{0.4pt}}

\vspace{1.9cm}

\noindent
{\bf $^{\dagger}$Corresponding author:} \vspace{-.15cm} \\ 
Andrew Berdahl, Santa Fe Institute, 1399 Hyde Park Rd,
Santa Fe, New Mexico, USA, 87501 \vspace{-.2cm} \\ 
\phone: (505) 946-2743 $\vert$ \Faxmachine: (505) 982-0565 $\vert$ \Email: berdahl@uw.edu  \\

\vspace{-.6cm}

\noindent
Submitted to \emph{Ecology Letters}.

\newpage

\section*{INTRODUCTION}

In a variety of ecosystems, a small change in a driving parameter can result
in a dramatic shift between disparate regimes.  Prominent examples
include the eutrophication of lakes
\citep{7eda3b0ae49d4888bba857664f4b460f,Smith2009201}, the degradation of
coral reefs \citep{mumby2007thresholds}, and the collapse of
forests \citep{scheffer2001catastrophic,cochrane1999positive}.  In the
particular case of tropical rainforests, it is hypothesized that an abrupt
shift between grass- and forest-dominated regimes is due to a fire-mediated
bistability \citep{staver2011global,hirota2011global,dantas2016disturbance}.  This
recent work has demonstrated a bistability at the macroscale and highlights
the importance of fire as a feedback mechanism that leads to this
bistability.  However, vegetation-fire feedbacks are inherently local spatial
processes \citep{laurance1991predicting,laurance2002ecosystem}.  Incorporating
such local processes may help understand how the size and shape of individual
forest fragments determine patterns in, and the stability of, the composite
system at the regional scale.

In tropical regions, fires that start in grassland typically burn only the
edges of forested regions, but do not spread within the forest
\citep{uhl1990deforestation,laurance2002ecosystem,cochrane2002fire,hoffmann2012fuels,brinck2017high}.
This behavior starkly contrasts with wildfires in temperate forests, where
the fire spreads readily through the trees via local ignition of neighboring
trees \citep{romme1989historical,turner2005landscape}.  This characteristic
of fire propagation in the tropics suggests that the ratio of the perimeter
of a forested patch (trees adjacent to grassland) to its area (i.e., its
protected interior) determines the relative vulnerability of such a patch to
fire.  On the other hand, the perimeter also determines the potential for the
outward expansion of a forested patch due to new tree growth at the
perimeter.  Thus the stability of an individual forest patch must depend on
its \textit{shape} as a result of the interplay between recession due to
exposure to fire and expansion due to perimeter-mediated growth.

At a larger scale, in Neotropical systems, burning has led to large-scale
forest fragmentation so that areas that were once covered by forest
subsequently consist of disjoint forest patches that are interspersed with
grassy areas; we term this type of morphology as savanna
\citep{pueyo2010testing,cochrane2003fire,cochrane1999positive,nepstad1999large}. The
state of the system is then driven by the collective fate of its constituent
forest patches.  Hence, edge effects have the potential to drive the dynamics
of forest patches at all scales: from the collapse of individual patches to
large-scale regime shifts between forest and savanna.  While it has been
shown that harsher conditions at patch edges are deleterious to trees and
increase the trees' tendency to burn
\citep{laurance1991predicting,laurance2002ecosystem,uhl1990deforestation},
the way such edge effects scale up to determinate the shape and stability of
forest patches is unknown.

In this work, we investigate the mechanisms that govern the size and shapes
of forests in the Brazilian Cerrado, and attempt to understand how
perturbations in the shape of a forest can affect its stability.  To do so,
we first examine high-resolution satellite imagery ($\sim 30\times 30$m
resolution \citep{hansen2013high}) of savanna-forest ecotones in the Cerrado
to quantify the statistical properties of the forested areas.  We find a
robust scaling relation between the perimeter, $P$, and area, $A$, of forest
patches, in which $P\sim A^{\gamma}$ with $\gamma \approx 0.69$.

To explain these observations, we introduce a mechanistic forest-fire model
that is specifically tailored to account for the particular phenomenology of
fire propagation in tropical regions that consist of both forested and grassy
areas.  In our model, fires spread readily through grass but less readily
through forested areas.  We denote this heterogeneously burning grass and
trees process as the BGT model. The spatial version of the BGT model
reproduces the same perimeter to area scaling relationship that we observe in
the empirical data.  We also show that deviations from this area-perimeter
relation can be used to infer the stability of individual forest patches and
thus suggest what interventions might lead to, or prevent, forest
fragmentation and the collapse of forest cover. These local spatial features
appear remarkably robust to changes in model parameters.

Finally, we observe interesting macro-scale ($>>30$m) behavior in both
simulations of our spatial BGT model and numerical solutions of its
mean-field version.  Specifically, both models exhibit a cusp bifurcation
\citep{kuznetsov2013elements} in total tree cover, in which both
discontinuous (first-order) or continuous (second-order) transitions observed
between forested and savanna states as a function of model parameters. In the
former case, microscopic perturbations to parameters can lead to macroscopic
changes in the amount of forest cover.

\section*{MATERIAL AND METHODS}

\subsection*{Data Collection}

\begin{figure}
  \centering \includegraphics[width=\linewidth,angle=-0]{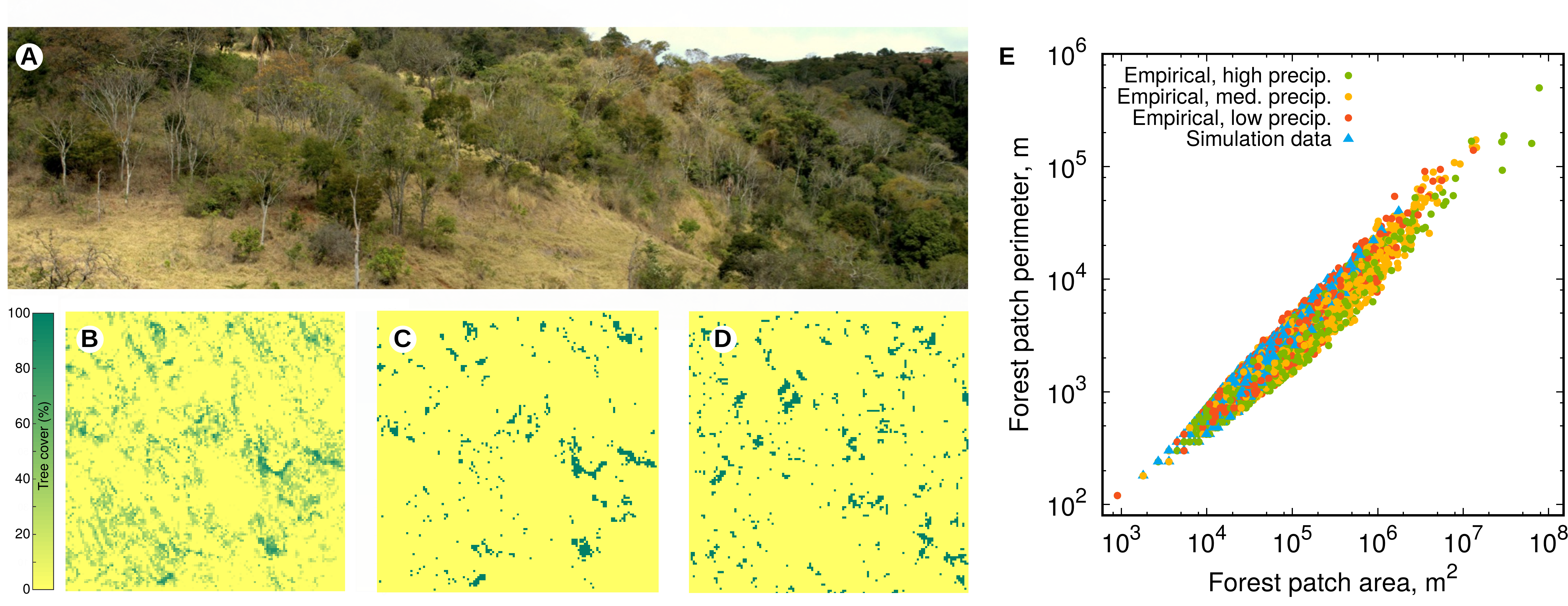}
  \caption{{\bf{Forest shapes.}}  {\bf{(A)}} Mixed forest-grassland ecosystem
    characteristic of the study region, the Brazilian Cerrado.  {\bf{(B)}}
    Raw forest cover percentage inferred from the high-resolution
    remote-sensing data, over a 4 km $\times$ 4 km area. {\bf{(C)}} Binary
    grass-forest state obtained using a $50\%$ threshold on the data of panel
    (B).  {\bf{(D)}} Snapshot of our spatial model, with same color scheme
    and spatial scale as (C). {\bf{(E)}} Perimeter versus area of individual
    forest patches. The circles denote empirical data (separated in three
    classes according to precipitation levels) while the triangles show one
    set of simulation data (see Table~\ref{tab:scaling} for parameter
    values).}
\label{fig1}
\end{figure}

The Brazilian Cerrado contains a mosaic of grassy areas and forests that are
interspersed throughout the landscape (Fig.~\ref{fig1}A-C).  We downloaded
tree cover data for the year 2000 (\cite{hansen2013high}, available from:\\
{http://earthenginepartners.appspot.com/science-2013-global-forest}) and
analyzed the subset of this data that corresponds to 86 Category-II protected
areas in the Brazilian Cerrado \citep{IUCN}, corresponding to a total area of
5.16 million hectares.  We focused on the Cerrado because of its mixed
forest-savanna ecotone and specifically IUCN-designated protected areas
within it to assess fire dynamics in forests and grasslands where there has
been no deliberate human intervention.  The resolution of the data is
$30\times 30$m and each cell is characterized by a continuous tree-cover
variable \citep{hansen2013high}, from which we assign each cell as either
forest when it contains $>50$\% tree cover or grass when it contains $<50$\%
tree cover.  The distribution of tree cover in these cells follows a roughly
bimodal distribution, with peaks at very high and very low tree cover, see
Appendix Fig.\ A.2.  Consequently, our definitions for forest and grass
patches are not sensitive to minor changes in the threshold used. As a check,
we performed a sensitivity analysis to our assumption of a 50\% threshold,
and replicated the analysis using a 75\% tree cover threshold. The results
are reported below and consistent with those found using the 50\% threshold
(Table~\ref{tab:scaling}).

We then combined forest cells that shared an edge into discrete forest
patches, and discarded any patches that were not fully contained within the
area of interest since the perimeters of these boundary patches would be
artificially short. From this data, we then calculated the area and perimeter
length of each forest patch.

\begin{landscape}
\begin{table}[htb]
\caption{Summary of scaling results for simulations and empirical data.}
\centering
\begin{tabular}{l l c c l l l l l l l c c l}

\toprule[0.5pt]

& Description &  Lattice size & Time step   & $p_{\lambda}$ & $p_{\alpha}$ & $p_{\beta}$ & $p_{f}$ & $p_{G}$ & $p_{T}$ & $p_{\mu}$  & Scaling & Confidence interval & Presented\\

\hline

 & low $\alpha/\beta$ ratio  & 768 & one year & 1.0 & 0.03 & 0.0003 & $6.2\cdot10^{-7}$ & 0.9 & 0.1 & 1.0 & 0.68947 &[0.68205 0.69698] & \\ 
 \parbox[t]{2mm}{\multirow{4}{*}{\rotatebox[origin=c]{90}{~~~~~~~~Simulations}}} &
 medium $\alpha/\beta$ ratio & 768 & one year & 1.0 & 0.03 & 0.001 & $7.0\cdot10^{-7}$ & 0.9 & 0.1 & 1.0 & 0.71111 & [0.70150 0.72083]  & Fig.~\ref{fig1}\\  
 & alternate lattice & 512 & one year & 1.0 & 0.03 & 0.001 & $1.6\cdot10^{-6}$ & 0.9 & 0.1 & 1.0 & 0.70690 & [0.69718 0.71672] & Fig.~\ref{fate} \\ 
 & alternate fire spread & 512 & one year & 1.0 & 0.03 & 0.0003 & $7.0\cdot10^{-7}$ & 0.8 & 0.2 & 1.0 & 0.68639 & [0.67746  0.69541] & \\ 
& alternate time step & 512 & 1/10 year & 0.1 & 0.004 & $10^{-5}$ & $4.0\cdot10^{-7}$ & 0.9 & 0.1 & 1.0 & 0.66622 & [0.66077 0.67170]  & \\ 
& high $\alpha/\beta$ ratio 1 & 256 & 1/10 year & 0.1 & 0.022 & $1.0\cdot10^{-6}$ & $3.4\cdot10^{-6}$ & 0.9 & 0.1 & 1.0 & 0.66914 & [0.64761 0.69130] & Appendix Fig.\ C4 \\ 
& high $\alpha/\beta$ ratio 2 & 256 & 1/10 year & 0.1 & 0.0022 & $2.0\cdot10^{-6}$ & $3.9\cdot10^{-6}$ & 0.9 & 0.1 & 1.0 & 0.69191 & [0.68228 0.70166] & \\

\hline

 & 50\% all data & -- & -- & -- & -- & -- & -- & -- & -- & -- & 0.69167 & [0.68234, 0.70111] & \\
 & 75\% all data & -- & -- & -- & -- & -- & -- & -- & -- & -- & 0.68382 & [0.67577, 0.69195] & \\
 \parbox[t]{2mm}{\multirow{6}{*}{\rotatebox[origin=c]{90}{~~~~~~~~Empirical}}} 
& 50\% low precip.\ & -- & -- & -- & -- & -- & -- & -- & -- & -- & 0.69630 & [0.67682, 0.71623] & Fig.~\ref{fig1} \\ 
  & 50\% med.\ precip.\ & -- & -- & -- & -- & -- & -- & -- & -- & -- & 0.70435 & [0.69262, 0.71625] & Fig.~\ref{fig1} \\ 
 & 50\% high precip.\ & -- & -- & -- & -- & -- & -- & -- & -- & -- & 0.67826 & [0.65163, 0.70585] & Fig.~\ref{fig1} \\ 
& 75\% low precip.\ & -- & -- & -- & -- & -- & -- & -- & -- & -- & 0.70748 & [0.69649, 0.71861] &  \\
  & 75\% med.\ precip.\ & -- & -- & -- & -- & -- & -- & -- & -- & -- & 0.70000 & [0.68601, 0.71424] &  \\ 
 & 75\% high precip.\ & -- & -- & -- & -- & -- & -- & -- & -- & -- & 0.65833 & [0.63144, 0.68620] &  \\

\bottomrule[0.5pt]
\end{tabular}
\label{tab:scaling}
\end{table}
\end{landscape}

\subsection*{Spatial BGT Model}

We are interested in understanding how the feedback between fire and the
regrowth of vegetation determines the distribution of forest shapes and how
the geometrical shape of a forest determines its stability.  To this end, we
formulate a spatial version of our BGT model that incorporates the basic
features that forest patches primarily grow outward at their perimeters and
burn exclusively at their edges.  The four fundamental aspects of our model
that parsimoniously account for tropical forest-grassland fire dynamics are:
(a) trees spread both locally at the perimeter and spontaneously due to seed
dispersal; (b) fires sporadically start in grassland but not in forests; (c)
fires percolate readily in grassland; (d) fires percolate only a short
distance into forested areas
\citep{laurance2002ecosystem,cochrane2002fire,hoffmann2012fuels}.  This model
is similar to, yet more parsimonious than, other cellular automata (CA)
forest-fire models designed for tropical forest-savanna systems
\citep{favier2004modelling,yassemi2008design,berjak2002improved}; however, it
contrasts with CA forest fire models from the physics literature
\citep{bak1990forest,chen1990deterministic,drossel1992self,0953-8984-8-37-004,grassberger2002critical}
in that it incorporates heterogeneous burning between grass and trees, and
both propagative and dispersed tree growth. These are the key features that
allow us to look at forest shapes in a novel way.

We discretize the land area as a two-dimensional lattice (with periodic
boundary conditions) where sites interact with their four nearest neighbors.
Each site can be occupied by trees, by grass, by ash/dirt, or by burning
material---either burning grass or trees.  At each time step, the densities
of tree, grass, ash, and burning sites, $T$, $G$, $A$, and $B$ evolve
according to the following processes:
\begin{enumerate}

\item[1.] Grass growth: An ash site turns to a grass site with probability
  $p_\lambda$.
  
\item[2.] Tree spreading: A tree site converts a \textit{neighboring} grass
  or ash site into another tree site with probability $p_\alpha$.

\item[3.] Spontaneous tree growth: Any grass or ash site can turn into a tree
  site with a small probability $p_\beta$.  Such a process arises from the
  long-distance dispersal of tree seeds.

\item[4.] Fire ignition and propagation: A grass site ignites with
  probability $p_f$.  Once a fire starts, it spreads as follows: (a) grass
  sites ignite with high probability $p_{\scriptscriptstyle G}$ while (b)
  tree sites ignite with small probability $p_{\scriptscriptstyle T}$, both
  from every neighboring burning site.  A burning site turns to ash after
  having the opportunity to spread to neighbouring sites.  We repeat this
  spreading process until no active fires remain. This means fires are
  `instantaneous' when compared to other timescales.

\end{enumerate}

The first three steps account, in a minimalist way, for vegetation growth,
while the final step accounts for consumption of vegetation by fire.  The
system is typically initialized in a state with no trees. Simulations begin
with a 500-time-step burn-in period, to reach the steady state, before we
start recording results, which is done only every 50 time steps to avoid
autocorrelations.

\subsubsection*{Parameter values}

\textbf{Timescale}: It is helpful define a time step to fix appropriate
parameter values. We choose an annual timescale, so that one time step of the
model corresponds to one year. To determined the sensitivity of our
simulation results to the choice of time step, we also simulate the case in
which a time step corresponds to $1/10$th of a year and adjust $p_\lambda$
such that our fastest rate, the rate of grass growth, remains the same.  This
sampling over different time scales allows us to test the impact, or lack
thereof in this case, of having either an instantaneous or probabilistic
grass regrowth.

\textbf{Grass growth}: Tropical grasses grow back roughly annually, and grass
fire intensity saturates after one year \citep{govender2006effect}.
Therefore, in simulations with an annual time step we use $p_\lambda=1$,
whereas we use $p_\lambda=0.1$ for a time step equal to a tenth of a year ---
both corresponding to grass growing back on an annual time scale.

\textbf{Forest propagation}: Based on tree maturation times on the order of
several decades
\citep{hoffmann2003comparative,rossatto2009differences,hoffmann2012ecological}
and our cell size of $30\times 30$ meters, we estimate that a forest
propagates at $\sim 1$m/year, and therefore set $p_{\beta}=0.03$, which is
also consistent with empirical observations \citep{durigan2006successional}
and previous modeling efforts \citep{favier2004modelling}.

\textbf{Forest dispersal}: Obtaining a quantitative estimate of the rate of
generation of new forest patches due to seed dispersal is more
difficult. Although long-distance dispersal is common
\citep{nathan2008mechanisms,nathan2006long,fragoso2003long}, the rate at
which new forest patches appear should be much lower than that of propagative
forest growth \citep{clark1999seed}. Because our model has no memory of how
long ago each patch was burned, the probability for creating a new forest
patch must incorporate not only the probability of seed dispersal, but also
the probability that such a patch avoids fire long enough for a $30\times 30$
meter stand of trees to mature and close a canopy. Therefore we set that rate
to be at least a factor of 30 lower than propagative forest growth.  However,
because of the uncertainty in this rate we explore a large range of values
for $p_{\beta}$ in our simulations.

\textbf{Fire ignition}: Typically, any given grassy area in this region burns
every 3--7 years \citep{oliveira2002cerrados,hoffmann2012ecological}.  We
therefore use $p_f \sim 1/(5L^2)$, which corresponds to a time for a fire to
re-occur at a given spatial location of approximately 5 years for a
simulation on a lattice of linear size $L$ and to a high fire spread
probability through grass ($p_G$).

\textbf{Fire propagation}: Because fires spread readily in grasslands but not
through tropical forests
\citep{uhl1990deforestation,laurance2002ecosystem,cochrane2002fire,hoffmann2012fuels,brinck2017high},
we use fire spread probabilities of $p_{\scriptscriptstyle G} = 0.8-0.9$ in
grass and $p_{\scriptscriptstyle T} = 0.1-0.2$ in trees. However our results
are robust as long as $p_{\scriptscriptstyle G}$ and
$p_{\scriptscriptstyle T}$ are respectively above and below the site
percolation threshold on the square lattice; that is,
$p_{\scriptscriptstyle G} > p_c = 0.5927\ldots$ and
$p_{\scriptscriptstyle T}<p_c$) \citep{aharony2003introduction}.  This means
that a grass fire can percolate throughout the system, while a tree fire is
necessarily finite in spatial extent.

\textbf{Spatial scale}: Consistent with our dataset, we assume each grid cell
corresponds to an area of $30\times 30$m. To check that our results are
insensitive to the lattice size, we simulated lattices from $256\times 256$
to $768\times 768$.

Please see Table~\ref{tab:notation} for a summary of parameter descriptions and values.

\begin{landscape}
\begin{table}[htb]
  \caption{Parameter descriptions and values used for simulations. Note that
    the numerical between the mean-field and spatial version of the model
    should in most cases not be directly compared. For example, values of
    $p_\beta$ and $\beta$ (and $p_f$ and $f$) do not directly correspond as
    the impact of the former depends on system size while the latter is
    defined in an infinite mean-field system.}
\centering
\begin{tabular}{c l l l l}
\cline{2-5}

\multicolumn{1}{l}{{}} &
\multicolumn{1}{l}{{}} &
\multicolumn{1}{l}{{}} &
\multicolumn{1}{l}{{}} &
\llap{\kern\tabcolsep\makebox[0pt]{Value}} \\
  
  \cmidrule(r){4-5}      
      
      \multicolumn{1}{l}{{}} &
        \multicolumn{1}{l}{{Symbol}} &
 \multicolumn{1}{l}{{Description}} &
  \multicolumn{1}{c}{Time step} &
    \multicolumn{1}{c}{Time step} \\

      \multicolumn{1}{l}{{}} &
              \multicolumn{1}{l}{{}} &
 \multicolumn{1}{l}{{}} &
  \multicolumn{1}{c}{1 year} &
    \multicolumn{1}{c}{1/10 year}  \\
   
\cline{2-5}

& $p_\lambda$ & Probability per time step that grass grows on an ash site  & \multicolumn{1}{c}{{1.0}} & \multicolumn{1}{c}{{0.1}} \\

& $p_\alpha$ & Probability per time step that a tree site converts an \emph{adjacent} grass or ash site to a tree site  & \multicolumn{1}{c}{{0.03}} & \multicolumn{1}{c}{{0.0022--0.04}} \\

 \parbox[t]{2mm}{\multirow{3}{*}{\rotatebox[origin=c]{90}{~~~~Spatial}}} & $p_\beta$ & Probability per time step that an ash or grass site, which is not necessarily bordering a tree site, converts to a tree site  & \multicolumn{1}{c}{{0.001--0.0003}} & \multicolumn{1}{c}{{$10^{-6}$--$10^{-5}$}} \\

 & $p_f$ & Probability per time step that a grass site spontaneously catches fire & \multicolumn{1}{c}{{$6.2\cdot10^{-7}$--$1.6\cdot10^{-6}$}} &  \multicolumn{1}{c}{{$4.0\cdot10^{-7}$--$3.9\cdot10^{-6}$}}\\

& $p_G$ & Probability that a fire spreads to an adjacent grass site  & \multicolumn{1}{c}{{0.8--0.9}} & \multicolumn{1}{c}{{0.9}}  \\

& $p_T$ & Probability that a fire spreads to an adjacent tree site & \multicolumn{1}{c}{{0.1--0.2}} & \multicolumn{1}{c}{{0.1}}  \\


\cline{2-5}

& $\lambda$ & Rate at which ash area is converted to grass area  &  &  \llap{\kern\tabcolsep\makebox[0pt]{0.05}} \\

& $\alpha$ & Rate at which treed area expands  &  &  \llap{\kern\tabcolsep\makebox[0pt]{0.001}} \\

 \parbox[t]{2mm}{\multirow{3}{*}{\rotatebox[origin=c]{90}{~~~~~~Mean-field}}} & $\beta$ & Rate at which grass area spontaneously converts to tree area  &  & \llap{\kern\tabcolsep\makebox[0pt]{0--1}} \\

& $f$ & Rate at which fire spontaneously starts in grass &  & \llap{\kern\tabcolsep\makebox[0pt]{0--3}} \\

& $\rho_G$ & Rate at which burning area spreads into grass area  &  &  \llap{\kern\tabcolsep\makebox[0pt]{0.9}} \\

& $\rho_T$ & Rate at which burning area spreads into tree area  &  &  \llap{\kern\tabcolsep\makebox[0pt]{0.1}} \\

& $\mu$ & Rate at which burning area is converted into ash area  &  &  \llap{\kern\tabcolsep\makebox[0pt]{$10^6$}} \\

\cline{2-5}
\end{tabular}
\label{tab:notation}
\end{table}
\end{landscape}

\subsection*{Mean-field BGT Model}

To develop an analytical understanding of the nature of the savanna-forest
transition, we also formulate a mean-field version of our spatial BGT model.
In this mean-field version description, the only variables are the global
densities of grass, tree, burning and ash sites.  The spatial location of
sites is are ignored and each site is assumed to neighbor to every other
site.  The simplicity of the mean-field description allows us to show
analytically that the BGT model exhibits both a discontinuous and a
continuous transition between a stable fully forested state and a savanna
state when basic system parameters, such as fire rate $f$ are varied.  A
discontinuous transition exhibits a sudden macroscopic drop in tree cover
with a microscopic change in fire frequency, while the tree cover decreases
continuously from a fully-forested state at a continuous transition as a
function of model parameters.  These transitions mirror the macroscale
behavior observed in simulations of the spatial BGT model.

In the mean-field description, spatial degrees of freedom are neglected and
the only variables are $B$, $G$, $T$, and $A$, which now denote the global
densities of burning sites, grass, trees, and ash.  Following the elemental
steps 1--4 of the spatial version outlined above, we show in the appendix
that these densities evolve in time as follows:
\begin{equation}
\label{MRE}
\begin{split}
\dot{B}&= fG +4\rho{\scriptscriptstyle G} BG + 4\rho_{\scriptscriptstyle
T}BT-\mu B\,,\\
\dot{G} &= \lambda A - \left(\beta+f\right)G - 4\alpha T G  \ - 4\rho_{\scriptscriptstyle G} B G \,, \\
\dot{T} &= 4\alpha T \left(G +A \right) + \beta G - 4\rho_{\scriptscriptstyle
  T} B T \, ,\\
\dot{A} &= \dot{B} - \dot{G} - \dot{T},
\end{split}
\end{equation}
where the overdot denotes the time derivative and conservation of the total
area mandates that $\dot B+\dot G+\dot T+\dot A=0$.  Here, the rates
($\lambda,\alpha,\beta,f,\rho_{\scriptscriptstyle G},\rho_{\scriptscriptstyle
  T}$) correspond to the probabilities that were defined in the 4 steps of
the spatial BGT model (see Table~\ref{tab:notation} for parameter definitions
and values). An additional rate---the rate $\mu$ at which burning sites turn
to ash---is needed because the continuous-time dynamics of the mean-field
description does not allow for instantaneous events.  Instead, the rate at
which burning material turns to ash is taken to be arbitrarily large so that
it exceeds all other rates in the system (Table~\ref{tab:notation}).

\section*{RESULTS}

\subsection*{Empirical} 

While the shapes of forest patches can be arbitrarily complicated, there are
two clear systematic and quantifiable features of the data.  The first is the
existence of a scaling relation between the perimeter $P$ and area $A$ of
forest patches. The data for the Cerrado region gives $P\sim A^{\gamma}$ with
$\gamma \approx 0.69$ (95\% confidence interval [0.68, 0.70]). Similar
scaling behavior arises for separate datasets that are partitioned into
regions of low (0--1100mm), medium (1100--1800mm), and high annual
precipitation ($>$1800mm) (Fig.~\ref{fig1}E, Table \ref{tab:scaling}) where
we find $\gamma \approx 0.70$ for the three precipitation cases.  Note that
for a compact shape, such as a circle or a square, $\gamma = 0.5$, while for
a dendritic shape (in the extreme case, a line) $\gamma \approx 1$.  The data
indicate that forest patches have a perimeter to area scaling behavior, with
$\gamma \approx 0.69$, consistently over several orders of magnitude in area.
This scaling law also arises in our spatial BGT model (to be discussed in the
next section).

\subsection*{Perimeter to area scaling in the spatial BGT model}

Simulations of the spatial BGT model produce a scaling relationship between
forest patch perimeters $P$ and their areas $A$ that is close to what is
observed in the Cerrado (Fig.~\ref{fig1}E). Using our most realistic
parameter values (upper four rows of Table~\ref{tab:scaling}), we observe
$P\sim A^\gamma$, with an average value of $\gamma \approx 0.70$ (95\%
confidence interval [0.69, 0.71]).  Simulations using a wider still range of
parameter values and environmental conditions demonstrate the robustness of
the scaling relationship: confidence intervals overlap regardless of the
annual rainfall in the Cerrado, and regardless of the model parameters in our
simulations as long as they lie within a large but realistic range of values
(Table~\ref{tab:scaling}).  In fact, parameter values for fire probability
and the relative recruitment rates of trees can be uncertain, and were
explored by varying their values by orders of magnitude in our
simulations. The consistency between the data and our model prediction for
the perimeter-area scaling gives credence to the mechanisms underlying the
BGT model. This correspondence between data and theory also allows us to use
the BGT model to make non-trivial predictions that are independent of
specific model parameter values, such as predicting the fate of individual
forest patches, as we show in the next section.

\subsection*{Fate of forest patches in the spatial BGT model}


\begin{figure}[bh!]
\centering
\begin{tabular}{c}
\includegraphics[width=0.7\linewidth]{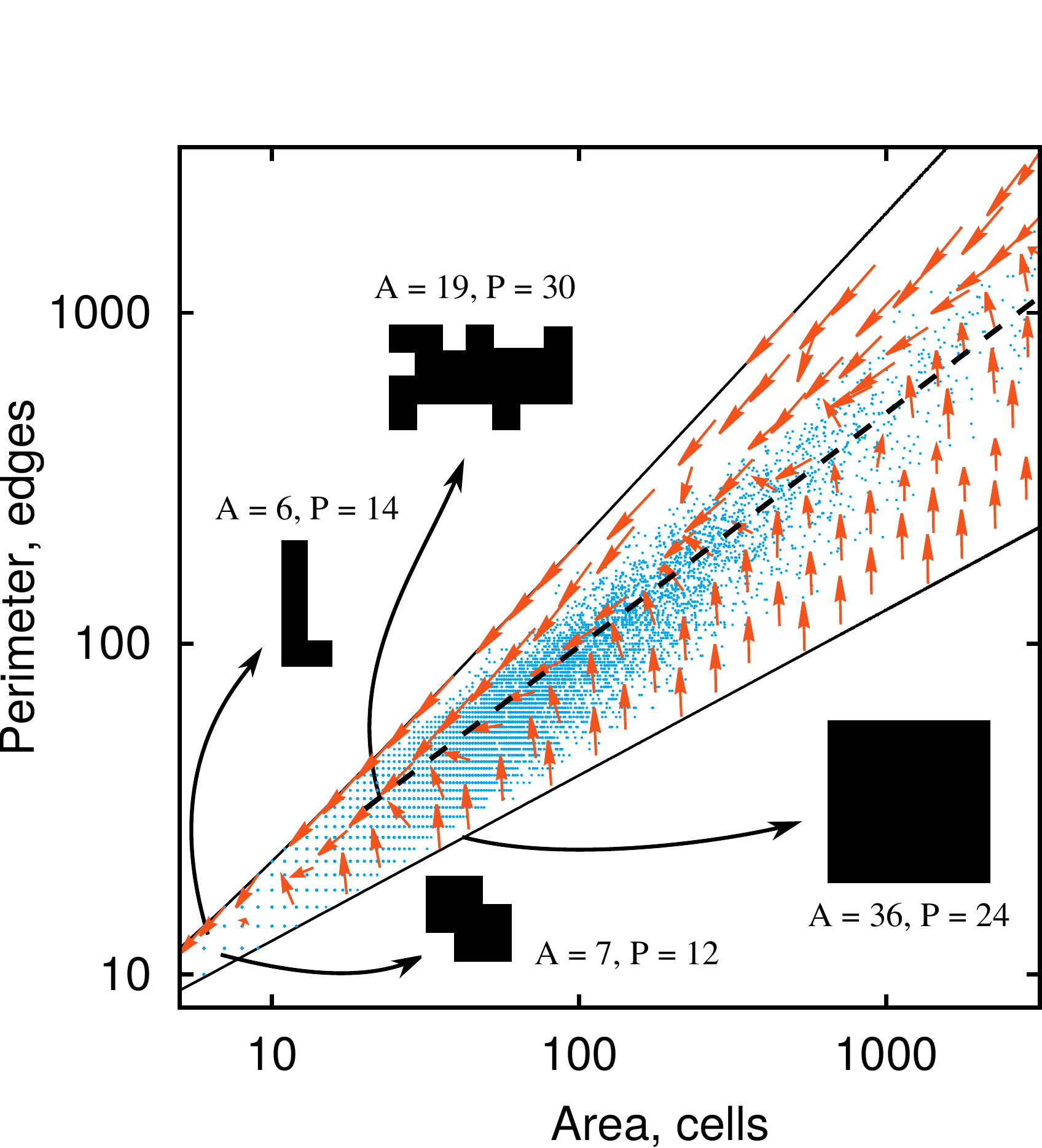}
\end{tabular}
\caption{\footnotesize{{\bf{Fate of forest patches.}} Evolution of forest
    patches of a given area and perimeter over 5 time steps (red arrows)
    using the ``alternate lattice'' set of parameters from Table
    \ref{tab:scaling}. The 5 time steps therefore provide the expected change
    in shape over a 5 year period.  When a forest is fragmented by fire, we
    track the largest remaining connected patch.  The blue data points
    correspond to the empirical data.  The allowable cone corresponds to all
    possible forest shapes---dendritic shapes at the upper limit, where the
    perimeter $P$ is proportional to the area $A$, and compact shapes at the
    lower limit, where $P\sim A^{1/2}$.  Some representative starting
    configurations and their corresponding locations in the perimeter-area
    plane are shown.  The dashed curve $P\sim A^{0.7}$ corresponds to the
    locus where the patch evolution is visually the slowest.}\label{fate}}
\end{figure}%

The results from our model suggest that forest patches converge on an
steady-state shape as a result of the balance between expansion by tree
propagation and their erosion by fire. We numerically test this hypothesis by
simulating the dynamics of individual patches of diverse sizes and shapes in
a system that have otherwise reached a steady-state.  For this test, we
artificially create a forest patch with a given shape and embed it in a much
larger steady-state mosaic of grassy and forested areas. The size of this
artificial patch ranges from 4 cells up to $10^4$ cells.  From initially
rectangular patches with aspect ratios in the range of 1:1 to 1:150, we
randomly change a fraction of the boundary tree sites to grass sites, in such
a way that the overall patch remains connected and also that most of the
allowable range of perimeter-area space is sampled (Fig.~\ref{fate}).  This
allowable range spans the physically realizable portion of the perimeter-area
plane where $P\sim A^\gamma$, with $1/2\leq \gamma\leq 1$.  A few examples of
these initial forest patches of small sizes are shown in the periphery of
Fig.~\ref{fate}.  We simulated at least $10^5$ distinct initial forest
patches for a given set of parameter values and ran each initial state for
five years (5 or 50 time steps (Figs.~\ref{fate} and C.4 respectively),
depending on the timescale).  The arrows in Fig.~\ref{fate} indicate the
magnitude and direction of the change in the area and perimeter from the
initial state, averaged over many realizations. If a forest patch is
fragmented by fire, we track the largest remaining fragment of the original
forest, this fragmentation mostly occurs in the top half of the allowable
shape range (above the dashed line in Fig.~\ref{fate}). In the bottom half,
most forest patches grow in area, but this growth is much slower than the
decrease observed in forests that shrink due to fire. Altogether, this
experiment allows us to better understand the impact of the shape of a
specific forest patch on its stability.
  
\subsection*{Phase transitions in the spatial BGT model}

\begin{figure}[h!]
\centering
\begin{tabular}{c}
\includegraphics[width=0.5\linewidth]{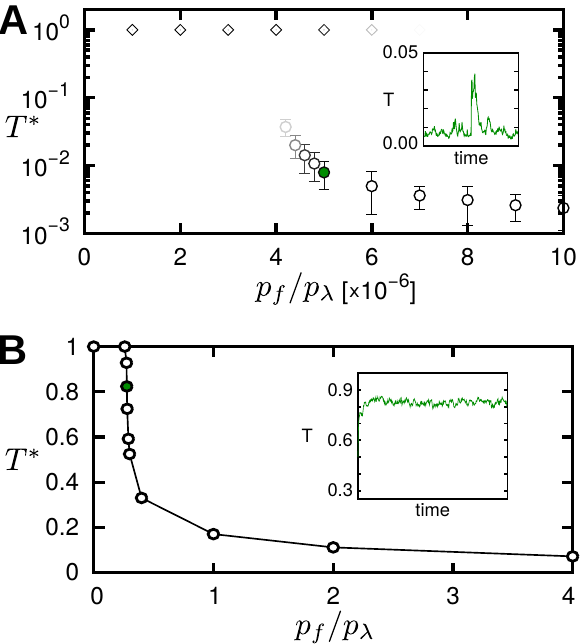}
\end{tabular}
\caption{\footnotesize{{\bf{Phase transitions in the spatial BGT model.}}
    {\bf{(A)}} Discontinuous transition for small spontaneous tree growth,
    $p_\beta$. The circles and squares indicate the steady-state values of
    forest cover, $T^*$, as the fire probability, $p_f$, is varied.  Their
    opacity is related to the probability of reaching this steady state from
    random initial conditions $T(0)=0.1$ (circles) and $T(0)=0.2$ (squares).
    Simulations are on a 512$\times$512 square lattice with
    $p_\beta=2\times 10^{-5}$ and $p_\alpha = 4\times 10^{-3}$.  {\bf{(B)}}
    Continuous transition using no propagative tree spreading
    ($p_\alpha = 0$), $p_\beta = 0.02$, and $p_\lambda=0.1$ on a
    256$\times$256 square lattice.  Standard deviations are smaller than
    marker size.  The insets shows total tree cover as a function of time
    ($T(t)$) to illustrate the behaviour of the system at equilibrium. We
    observe large stochastic fluctuations in tree coverage over time close to
    the discontinuous transition (panel A) that are not present in the
    discontinuous case (panel B). The filled green data points in the main
    figures corresponds to the set of parameters used to produce the time
    series shown in the insets.  }\label{transitions}}
\end{figure}

The state $T=1$ is an absorbing state of our model; once trees have invaded
the system, no more dynamics can occur. In our simulations, we therefore
always initiated the system at $T=0$ to maximize the probability of reaching
the savanna steady state and minimize the probability of reaching the
absorbing state.  For example, for an initial forest cover of $T=0.2$ and
fire probability $p_f/p_{\lambda} = 5\times 10^{-6}$ (corresponding to the
middle of the bistable regime in Fig.~\ref{transitions}A), a steady savanna
state is reached approximately only 10\% of the time while the remaining 90\%
of simulations reach the absorbing state.

This sensitivity to the initial condition indicates that the system is close
to a separatrix in parameter space, where small changes in the initial
densities can lead to the system being driven towards different basins of
attractions.  Correspondingly, there is a discontinuous change in the steady
state of the system and a concomitant hysteresis for small changes in the
initial condition.  These behaviors are consistent with the bistability
hypothesized to exist in tropical forests and savannas
\citep{staver2011global,hirota2011global} and predicted by related models of
mixtures of grass and forest
\citep{schertzer2015implications,durrett2015coexistence}.

The phenomenology that we predict changes drastically when spontaneous tree
growth due to seed dispersion is the dominant contributor to forest growth, a
factor that is often omitted in previous models.  As illustrated in
Fig.~\ref{transitions}B, we set an extreme value of $p_\alpha = 0$ and
examine the behavior for a large value of $p_\beta $ ($10^3$ times larger
than for the data in Fig.~\ref{transitions}A) that promotes spontaneous tree
growth on the grassy areas.  In this extreme situation, there is a continuous
transition (no discrete jump in tree cover) between the savanna and the fully
forested state, as well as much smaller fluctuations in the time series of
total amount of observed tree cover (Fig.~\ref{transitions}B).  This scenario
with extremely large $\beta$ leads to an unrealistically rapid growth of
trees, but it also illustrates the importance of the $\alpha/\beta$ ratio in
shaping the phenomenology of our model. This feature of the BGT model is
better studied in its mean-field version.

\begin{figure}[bh!]
\centering
\begin{tabular}{c}
\includegraphics[width=\linewidth]{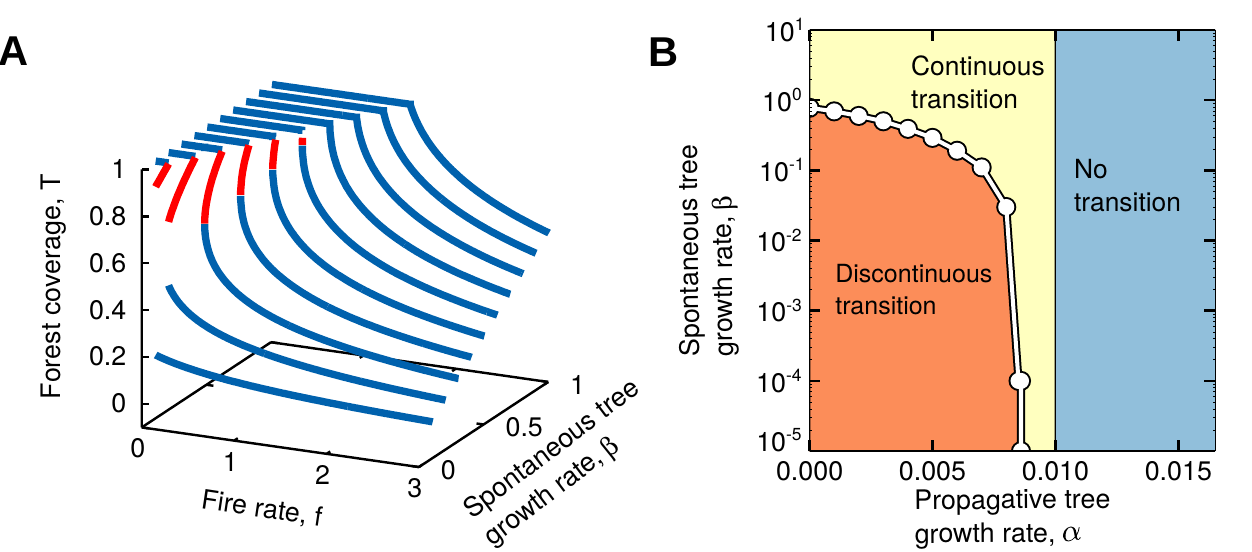}
\end{tabular}
\caption{\footnotesize{{\bf{Phase transitions in mean-field BGT model.}}
    {\bf{(A)}} Solution of the mean-field model that illustrates the change
    between a discontinuous and continuous transition. Blue curves are stable
    steady states, whereas red curves are unstable solutions (missing
    segments go to the unphysical regime $f<0$). Note that all the
    interesting dynamics occur in the realistic regime of slow fire rates $f$
    and spontaneous tree growth $\beta$. {\bf{(B)}} Nature of the phase
    transition from the fully treed state ($T^* = 1$) to a mixed state
    ($0 < T^* < 1$) as a function of propagative and spontaneous tree growth
    ($\alpha$ and $\beta$ respectively). White markers were obtained by
    solving for the critical values of $\beta_c$ that separate the
    discontinuous and continuous transition regimes. The value of $\alpha_c$
    that marks the border of the regime without transition can be derived
    using the Jacobian of Eqs. \eqref{MRE} (see Appendix). Note that the
    numerical values of $p_\beta$ and $\beta$ (and $p_f$ and $f$) do not
    directly correspond as the impact of the former depends on system size
    while the latter is defined in an infinite mean-field system. The large
    range of $\beta$ is shown for completeness, but in reality we would not
    expect $\beta$ to be larger than $\alpha$ (uppermost region of
    figure). In fact, the the rate of propagative growth $\alpha$ is by far
    the most critical mechanism discerning the regime
    shifts. }\label{transitionsMF}}
\end{figure}%

\subsection*{Phase transitions in the mean-field BGT model}

We can solve for the steady states of the rate equations \eqref{MRE}
numerically and gain insights into the nature of the observed phase
transitions. These solutions give rise to a cusp bifurcation
\citep{kuznetsov2013elements}, leading to three distinct regimes in parameter
space: (i) a regime with a single stable solution at a fully forested state
$T^*=1$; (ii) a regime with a single stable solution which is a mixed-state
that corresponds to a savanna ($0<T^*<1$); (iii) a regime which has two
stable solutions, $T^*_{(1)} = 1$ and $0<T^*_{(2)} < 1$, which are absorbing
states for different sets of initial conditions that are demarcated by a
separatrix.  The transitions between the fully forested and the savanna state
are particularly interesting.  For sufficiently large values of $\beta$ ---
i.e., large rate of seed dispersal and ensuing forest patch generation ---
there is a continuous transition between the forest-dominated state ($T=1$)
and the savanna state ($T<1$) as a function of the fire rate $f$
(Fig.~\ref{transitionsMF}A).  However, for $\beta$ less than a critical
value, $\beta_c$, which can only be determined numerically, the tree cover,
$T$, undergoes a discontinuous jump transition instead.  This discontinuity
is accompanied by both bistability and hysteresis (Fig.~\ref{transitionsMF}A
and Appendix Fig.~E.6).  Thus the existence of bistability is determined in
part by which tree-growth mechanism---spontaneous growth ($\beta$) or
neighbor-driven growth ($\alpha$)---is more effective in creating additional
trees.

\section*{DISCUSSION}

Fire spread is an inherently local process, yet is thought to drive
vegetation dynamics at a regional scale
\citep{hirota2011global,higgins2012atmospheric}.  Localized empirical studies
have provided strong support for the hypothesis that different rates of fire
spread in savannas and forests \citep{hoffmann2012fuels} largely determine
the growth and mortality of tree species \citep{hoffmann2009tree}, which in
turn can stabilize the savanna-forest boundary
\citep{hoffmann2012ecological}.  By modeling the different fire propagation
probabilities in savannas and forests, we are able to reproduce empirically
observed patterns at both local scales (the shape and distribution of forest
patches) and regional scales (the distribution, or bimodality, of forest
coverage).  Thus, we connect local mechanisms of fire dynamics, which operate
on the scale of meters, to regional patterns of forest and grassland
distribution, which operate on the scale of hundreds of kilometers.

Using our spatial model, we showed that commonly observed ``edge effects''
\citep{brinck2017high,laurance2002ecosystem,cochrane2002fire,uhl1990deforestation}
drive the stability and fate of individual forest patches.  Importantly, the
two competing processes of exposure to fire and propagative growth do not
balance because of their very different time scales. For a forest of a given
size, having a large perimeter means that it is more likely to be exposed to
fire before it receives any benefit from increased tree cover due to
propagative growth. Contrarily, if too dense, a forest will not grow
sufficiently quickly to offset its exposure to fire, which will lead to an
increase in perimeter by encroachment of grassland at the forest edges. The
forests that do survive are those that end up with intermediate shapes to
ensure enough perimeter growth to offset their exposure to
fire.

At the scale of individual forest patches, the basic questions that we
addressed were: Will a given forest patch become denser or sparser?  Will it
grow or will it shrink?  Dense shapes (e.g., circular or square forest
patches), which correspond to the lower limit of the allowable range in
Fig.~\ref{fate}, tend to grow in our simulations, while dendritic shapes that
correspond to the upper limit of the range, tend to shrink and/or fragment.
A critical feature of this plot is that the locus of points where the
observed steady-state perimeter-area relation $P\sim A^\gamma$ holds, also
corresponds to the locus where the change in the forest perimeter and area is
visually the smallest.  Thus the most stable forest patches in the spatial
BGT model are those that conform to the perimeter-area relation observed in
the Brazilian Cerrado (Fig.~\ref{fig1}).  These results are virtually
independent of our choice of parameters. Only extremely unrealistic regimes
of $\beta > \alpha$, i.e. with faster dispersed growth than propagative
growth, seem to produce patches with a linear perimeter-area scaling that
deviate from the empirical data (see Appendices).

Our spatial model could be a useful tool for resource managers and other
stakeholders of tropical forest-savanna systems, because of its potential to
predict the fate and risk of collapse of individual forest fragments based on
their area and perimeter. This is particularly so due to our model's lack of
sensitivity to the specific parameter values. For example, our model could be
used to predict how a specific deformation, such as logging, would affect the
stability of a forest patch.  However, it is also important to keep in mind
that forest edges play a role beyond forest stability. For example, edges are
critical to maintaining biodiversity \citep{turner1996species}.  Nonetheless,
the validation of our predictions for the fate of specific forest patches,
using remotely sensed data with high temporal resolution, will aid in these
quantitative predictions.

At the regional scale, we found that our spatial model can exhibit either a
sudden discontinuous transition, or a steep but smooth continuous transition,
from a fully forested state to a savanna ecotone when parameters change.  At
this larger scale, the parameterization profoundly affects the results, in
contrast, the observations for the fates of individual patches, which were
insensitive to specific parameters.  Studying the mean-field version of our
model allowed us to show that this parameter sensitivity is the result of a
cusp bifurcation as a function of three basic parameters---the fire rate $f$,
the tree propagation rate $\alpha$, and the spontaneous tree growth rate
$\beta$.  The nature of the transition is set by whether tree growth is
dominated by propagative forest expansion, which leads to a discontinuous
transition, or by tree growth due to seed dispersal, which leads to a
continuous transition.  This means that factors to increase the recruitment
of new forest patches could move the system into a regime with a continuous
transition.  Such an intervention would be desirable because in such a regime
there is less chance of runaway shrinking of forest cover than in the regime
with a discontinuous transition.  Similarly, animal species that contribute
to long-distance seed dispersal \citep{nathan2008mechanisms}, especially into
grasslands, may play a critical role in limiting discontinuous (or critical)
regime shifts.

In previous work, bimodality in the empirical distributions of regional
forest cover is often used to hypothesize bistability, and thus discontinuous
transitions (criticality), in tropical forests.  Our BGT model is consistent
with that interpretation since it can also exhibit a discontinuous
transition. However, a continuous transition, if sufficiently steep, could
give rise to bimodality in forest cover (Fig.~\ref{transitions}B).  Thus, the
empirically observed bimodality is consistent even with our continuous
transition, so this bimodality is not necessarily evidence for bistability.
This critical distinction between our model and empirical interpretations has
implications for how sensitive tropical savanna-forest biomes will be to
global change (e.g., \cite{moncrieff2014increasing}).  Empirical studies that
quantify the response of various parameters in our model, such as the rate of
fire spread, the rate of local tree propagation, etc., to different climatic
regimes will allow us to predict the nature of the potential transitions
(continuous or discontinuous) between forested and savanna states.

Our parsimonious model comes with several caveats and assumptions.  Strictly
speaking, we have assumed that mortality when tree sites burn is $>50$\%, so
that even sites at 100\% tree cover will fall below the threshold following a
fire.  In reality, tree mortality due to fire can take on a wide range of
values, but our model indirectly accounts for this in two ways. First, a
higher value of tree growth due to spreading ($p_\alpha$) could capture the
fast regrowth of peripheral trees that were burned but not completely
killed. For this reason, some of our simulations use high $p_\alpha$ values
(`high $\alpha/\beta$ ratio 1' in Table~\ref{tab:scaling} corresponds to a
4.5 year time scale for tree spreading).  Two, one can think of $p_\alpha$ as
the product of the probability the trees will burn and the probability they
will then die. In this sense, when we varied $p_\alpha$ we effectively varied
the fire mortality, and found no change in results. Nonetheless, future
models could explicitly consider variability in tree mortality due to factors
related to microclimate that influence fire spread and intensity, as well as
community composition \citep{hoffmann2009tree} which can vary substantially
across the landscape \citep{pellegrini2016shifts}.

We have also assumed spatial homogeneity in our model.  For instance, all
grassy areas have an equal probability of burning, which would be the case if
there was random ignition across the entire area. In other ecosystems such as
coniferous forests, it has been documented that the spread of a single fire
can be quite heterogeneous due to a number of factors
\citep{turner2010disturbance}. In grasslands, which experience a higher
frequency of burning, we expect the heterogeneity in the effect of a single
fire ignition and spread to be reduced when multiple fire cycles are
considered, because it allows for a greater likelihood that any particular
location is burned. On the other hand, local factors, such as underlying
topography or even termite mounds \citep{levick2010regional}, have the
potential to influence fire spread. In these cases, spatial patterns may
arise due to the distribution of edaphic features. Similarly, heterogeneity
in fire ignition and spread in tropical forests can also be caused by their
proximity to roads \citep{nepstad2001road}.  Moreover, tree growth (other
than edge vs.\ random) is homogeneous in our model, but in the real world may
depend on geographic features, such a topology and proximity to
water. Indeed, in the empirical data we observe forest patches whose shape is
almost certainly governed by geography (Appendix Fig.\ A.3). However, while
these forests are large in area, they do not contribute many forest patches
(i.e., data points) in terms of numbers, so they do not dominate the
analysis.

We have assumed binary levels of tree cover (a site is either treed or grass)
and homogeneity in their composition. Trees, especially fire adapted ones
\citep{hoffmann2003comparative,rossatto2009differences}, may be sparsely
distributed through grassy areas.  Our model is consistent with this picture,
provided these trees do not significantly impede fire spreading through grass
(or if they do it can be incorporated into the fire spread rate $p_G$).
Certainly there is a wide diversity of tree types and demographic
characteristics of plants could influence processes such as tree growth and
fire spread. However the edge effects that we model are general across
species and communities in tropical forest-savanna biomes
\citep{brinck2017high,laurance2002ecosystem,cochrane2002fire,uhl1990deforestation}. Thus
a single tree type (or a homogeneous mix) as modeled here seems adequate to
address the questions at hand, but adding more detail (e.g.,
\cite{favier2004modelling,yassemi2008design,berjak2002improved}) may be
important for other questions.

Finally, testing the generality of our model in different savanna-forest
ecotones outside of South America will provide insight into its generality.
In African savanna, we may expect fundamentally different dynamics because of
the large role that elephants play in determining the distribution of
savannas and forests \citep{asner2016ecosystem,asner2009large}. Indeed, both
elephants and fire are hypothesized to act in concert to determine
alternative stable states of savanna and forest, but unclear whether such
edge effects emerge given the ability of elephants to topple large trees
\citep{dublin1990elephants}. In Australia, tree cover can be sensitive to
changes in precipitation even in relatively wet conditions.

In conclusion, regardless of model assumptions, we have shown that the low
spread rate of fires in forests and its high spread rate in savannas plays a
pervasive role in structuring the shape of forest patches across a
Neotropical savanna-forest landscape. The ability to infer the stability of
an ecosystem from the growth, recruitment, and fire-driven mortality allows
us to assess the persistence of both savanna and forest landscapes. Moreover,
the ability to infer stability from the shape of forest fragments presents a
potential tool for large-scale and rapid assessment of the stability of
tropical terrestrial ecosystems.  Understanding the key mechanisms at play in
shaping forests and their stability will aid in predicting the future of land
carbon sinks \citep{brinck2017high,pan2011large} and biodiversity
\citep{tilman1994habitat,alroy2017effects}.

\section*{ACKNOWLEDGMENTS}
We thank William F.\ Laurance, Munik Shrestha and two anonymous reviewers for
helpful suggestions.  This work has been supported by the Santa Fe Institute,
a James S. McDonnell Foundation Postdoctoral Fellowship (LHD), a Santa Fe
Institute Omidyar Postdoctoral Fellowship (AMB), a NOAA Climate and Global
Change Postdoctoral Fellowship (AFAP) and by the grants DMR-1608211 and
1623243 from the National Science Foundation (UB and SR), the John Templeton
Foundation (SR, AMB), and Grant No.\ 2012145 from the United States Israel
Binational Science Foundation (UB).


\section*{REFERENCES}

\begin{thebibliography}{}

\bibitem[IUC, 2016]{IUCN}
 (2016).
\newblock Category-{II}: National park, as defined by the {International Union
  for Conservation of Nature} {(IUCN)}.

\bibitem[Aharony and Stauffer, 2003]{aharony2003introduction}
Aharony, A. and Stauffer, D. (2003).
\newblock {\em Introduction to percolation theory}.
\newblock Taylor \& Francis.

\bibitem[Alroy, 2017]{alroy2017effects}
Alroy, J. (2017).
\newblock Effects of habitat disturbance on tropical forest biodiversity.
\newblock {\em Proceedings of the National Academy of Sciences},
  114(23):201611855.

\bibitem[Asner et~al., 2009]{asner2009large}
Asner, G.~P., Levick, S.~R., Kennedy-Bowdoin, T., Knapp, D.~E., Emerson, R.,
  Jacobson, J., Colgan, M.~S., and Martin, R.~E. (2009).
\newblock Large-scale impacts of herbivores on the structural diversity of
  african savannas.
\newblock {\em Proceedings of the National Academy of Sciences},
  106(12):4947--4952.

\bibitem[Asner et~al., 2016]{asner2016ecosystem}
Asner, G.~P., Vaughn, N., Smit, I.~P., and Levick, S. (2016).
\newblock Ecosystem-scale effects of megafauna in african savannas.
\newblock {\em Ecography}, 39(2):240--252.

\bibitem[Bak et~al., 1990]{bak1990forest}
Bak, P., Chen, K., and Tang, C. (1990).
\newblock A forest-fire model and some thoughts on turbulence.
\newblock {\em Physics Letters A}, 147(5):297--300.

\bibitem[Berjak and Hearne, 2002]{berjak2002improved}
Berjak, S.~G. and Hearne, J.~W. (2002).
\newblock An improved cellular automaton model for simulating fire in a
  spatially heterogeneous savanna system.
\newblock {\em Ecological modelling}, 148(2):133--151.

\bibitem[Brinck et~al., 2017]{brinck2017high}
Brinck, K., Fischer, R., Groeneveld, J., Lehmann, S., De~Paula, M.~D.,
  P{\"u}tz, S., Sexton, J.~O., Song, D., and Huth, A. (2017).
\newblock High resolution analysis of tropical forest fragmentation and its
  impact on the global carbon cycle.
\newblock {\em Nature Communications}, 8.

\bibitem[Chen et~al., 1990]{chen1990deterministic}
Chen, K., Bak, P., and Jensen, M.~H. (1990).
\newblock A deterministic critical forest fire model.
\newblock {\em Physics Letters A}, 149(4):207--210.

\bibitem[Clar et~al., 1996]{0953-8984-8-37-004}
Clar, S., Drossel, B., and Schwabl, F. (1996).
\newblock Forest fires and other examples of self-organized criticality.
\newblock {\em Journal of Physics: Condensed Matter}, 8(37):6803.

\bibitem[Clark et~al., 1999]{clark1999seed}
Clark, J.~S., Silman, M., Kern, R., Macklin, E., and HilleRisLambers, J.
  (1999).
\newblock Seed dispersal near and far: patterns across temperate and tropical
  forests.
\newblock {\em Ecology}, 80(5):1475--1494.

\bibitem[Cochrane, 2003]{cochrane2003fire}
Cochrane, M.~A. (2003).
\newblock Fire science for rainforests.
\newblock {\em Nature}, 421(6926):913--919.

\bibitem[Cochrane et~al., 1999]{cochrane1999positive}
Cochrane, M.~A., Alencar, A., Schulze, M.~D., Souza, C.~M., Nepstad, D.~C.,
  Lefebvre, P., and Davidson, E.~A. (1999).
\newblock Positive feedbacks in the fire dynamic of closed canopy tropical
  forests.
\newblock {\em Science}, 284(5421):1832--1835.

\bibitem[Cochrane and Laurance, 2002]{cochrane2002fire}
Cochrane, M.~A. and Laurance, W.~F. (2002).
\newblock Fire as a large-scale edge effect in amazonian forests.
\newblock {\em Journal of Tropical Ecology}, 18(03):311--325.

\bibitem[Conley et~al., 2009]{7eda3b0ae49d4888bba857664f4b460f}
Conley, D., Paerl, H., Howarth, R., Boesch, D., Seitzinger, S., Havens, K.,
  Lancelot, C., and Likens, G. (2009).
\newblock Ecology - controlling eutrophication: Nitrogen and phosphorus.
\newblock {\em Science}, 323(5917):1014--1015.

\bibitem[Dantas et~al., 2016]{dantas2016disturbance}
Dantas, V. d.~L., Hirota, M., Oliveira, R.~S., and Pausas, J.~G. (2016).
\newblock Disturbance maintains alternative biome states.
\newblock {\em Ecology Letters}, 19(1):12--19.

\bibitem[Drossel and Schwabl, 1992]{drossel1992self}
Drossel, B. and Schwabl, F. (1992).
\newblock Self-organized critical forest-fire model.
\newblock {\em Physical Review Letters}, 69(11):1629.

\bibitem[Dublin et~al., 1990]{dublin1990elephants}
Dublin, H.~T., Sinclair, A.~R., and McGlade, J. (1990).
\newblock Elephants and fire as causes of multiple stable states in the
  serengeti-mara woodlands.
\newblock {\em The Journal of Animal Ecology}, pages 1147--1164.

\bibitem[Durigan and Ratter, 2006]{durigan2006successional}
Durigan, G. and Ratter, J. (2006).
\newblock Successional changes in cerrado and cerrado/forest ecotonal
  vegetation in western sao paulo state, brazil, 1962--2000.
\newblock {\em Edinburgh Journal of Botany}, 63(1):119--130.

\bibitem[Durrett et~al., 2015]{durrett2015coexistence}
Durrett, R., Zhang, Y., et~al. (2015).
\newblock Coexistence of grass, saplings and trees in the {S}taver--{L}evin
  forest model.
\newblock {\em The Annals of Applied Probability}, 25(6):3434--3464.

\bibitem[Favier et~al., 2004]{favier2004modelling}
Favier, C., Chave, J., Fabing, A., Schwartz, D., and Dubois, M.~A. (2004).
\newblock Modelling forest--savanna mosaic dynamics in man-influenced
  environments: effects of fire, climate and soil heterogeneity.
\newblock {\em Ecological Modelling}, 171(1):85--102.

\bibitem[Fragoso et~al., 2003]{fragoso2003long}
Fragoso, J., Silvius, K.~M., and Correa, J.~A. (2003).
\newblock Long-distance seed dispersal by tapirs increases seed survival and
  aggregates tropical trees.
\newblock {\em Ecology}, 84(8):1998--2006.

\bibitem[Govender et~al., 2006]{govender2006effect}
Govender, N., Trollope, W.~S., and Van~Wilgen, B.~W. (2006).
\newblock The effect of fire season, fire frequency, rainfall and management on
  fire intensity in savanna vegetation in south africa.
\newblock {\em Journal of Applied Ecology}, 43(4):748--758.

\bibitem[Grassberger, 2002]{grassberger2002critical}
Grassberger, P. (2002).
\newblock Critical behaviour of the {D}rossel-{S}chwabl forest fire model.
\newblock {\em New Journal of Physics}, 4(1):17.

\bibitem[Hansen et~al., 2013]{hansen2013high}
Hansen, M.~C., Potapov, P.~V., Moore, R., Hancher, M., Turubanova, S.,
  Tyukavina, A., Thau, D., Stehman, S., Goetz, S., Loveland, T., et~al. (2013).
\newblock High-resolution global maps of 21st-century forest cover change.
\newblock {\em Science}, 342(6160):850--853.

\bibitem[Higgins and Scheiter, 2012]{higgins2012atmospheric}
Higgins, S.~I. and Scheiter, S. (2012).
\newblock Atmospheric co2 forces abrupt vegetation shifts locally, but not
  globally.
\newblock {\em Nature}, 488(7410):209--212.

\bibitem[Hirota et~al., 2011]{hirota2011global}
Hirota, M., Holmgren, M., Van~Nes, E.~H., and Scheffer, M. (2011).
\newblock Global resilience of tropical forest and savanna to critical
  transitions.
\newblock {\em Science}, 334(6053):232--235.

\bibitem[Hoffmann et~al., 2009]{hoffmann2009tree}
Hoffmann, W.~A., Adasme, R., Haridasan, M., T~de Carvalho, M., Geiger, E.~L.,
  Pereira, M.~A., Gotsch, S.~G., and Franco, A.~C. (2009).
\newblock Tree topkill, not mortality, governs the dynamics of savanna--forest
  boundaries under frequent fire in central brazil.
\newblock {\em Ecology}, 90(5):1326--1337.

\bibitem[Hoffmann et~al., 2012a]{hoffmann2012ecological}
Hoffmann, W.~A., Geiger, E.~L., Gotsch, S.~G., Rossatto, D.~R., Silva, L.~C.,
  Lau, O.~L., Haridasan, M., and Franco, A.~C. (2012a).
\newblock Ecological thresholds at the savanna-forest boundary: how plant
  traits, resources and fire govern the distribution of tropical biomes.
\newblock {\em Ecology letters}, 15(7):759--768.

\bibitem[Hoffmann et~al., 2012b]{hoffmann2012fuels}
Hoffmann, W.~A., Jaconis, S.~Y., Mckinley, K.~L., Geiger, E.~L., Gotsch, S.~G.,
  and Franco, A.~C. (2012b).
\newblock Fuels or microclimate? understanding the drivers of fire feedbacks at
  savanna--forest boundaries.
\newblock {\em Austral Ecology}, 37(6):634--643.

\bibitem[Hoffmann et~al., 2003]{hoffmann2003comparative}
Hoffmann, W.~A., Orthen, B., and Nascimento, P. K. V.~d. (2003).
\newblock Comparative fire ecology of tropical savanna and forest trees.
\newblock {\em Functional Ecology}, 17(6):720--726.

\bibitem[Kuznetsov, 2013]{kuznetsov2013elements}
Kuznetsov, Y.~A. (2013).
\newblock {\em Elements of applied bifurcation theory}, volume 112.
\newblock Springer Science \& Business Media.

\bibitem[Laurance et~al., 2002]{laurance2002ecosystem}
Laurance, W.~F., Lovejoy, T.~E., Vasconcelos, H.~L., Bruna, E.~M., Didham,
  R.~K., Stouffer, P.~C., Gascon, C., Bierregaard, R.~O., Laurance, S.~G., and
  Sampaio, E. (2002).
\newblock Ecosystem decay of amazonian forest fragments: a 22-year
  investigation.
\newblock {\em Conservation Biology}, 16(3):605--618.

\bibitem[Laurance and Yensen, 1991]{laurance1991predicting}
Laurance, W.~F. and Yensen, E. (1991).
\newblock Predicting the impacts of edge effects in fragmented habitats.
\newblock {\em Biological Conservation}, 55(1):77--92.

\bibitem[Levick et~al., 2010]{levick2010regional}
Levick, S.~R., Asner, G.~P., Chadwick, O.~A., Khomo, L.~M., Rogers, K.~H.,
  Hartshorn, A.~S., Kennedy-Bowdoin, T., and Knapp, D.~E. (2010).
\newblock Regional insight into savanna hydrogeomorphology from termite mounds.
\newblock {\em Nature communications}, 1:65.

\bibitem[Moncrieff et~al., 2014]{moncrieff2014increasing}
Moncrieff, G.~R., Scheiter, S., Bond, W.~J., and Higgins, S.~I. (2014).
\newblock Increasing atmospheric co2 overrides the historical legacy of
  multiple stable biome states in africa.
\newblock {\em New Phytologist}, 201(3):908--915.

\bibitem[Mumby et~al., 2007]{mumby2007thresholds}
Mumby, P.~J., Hastings, A., and Edwards, H.~J. (2007).
\newblock Thresholds and the resilience of caribbean coral reefs.
\newblock {\em Nature}, 450(7166):98--101.

\bibitem[Nathan, 2006]{nathan2006long}
Nathan, R. (2006).
\newblock Long-distance dispersal of plants.
\newblock {\em Science}, 313(5788):786--788.

\bibitem[Nathan et~al., 2008]{nathan2008mechanisms}
Nathan, R., Schurr, F.~M., Spiegel, O., Steinitz, O., Trakhtenbrot, A., and
  Tsoar, A. (2008).
\newblock Mechanisms of long-distance seed dispersal.
\newblock {\em Trends in ecology \& evolution}, 23(11):638--647.

\bibitem[Nepstad et~al., 2001]{nepstad2001road}
Nepstad, D., Carvalho, G., Barros, A.~C., Alencar, A., Capobianco, J.~P.,
  Bishop, J., Moutinho, P., Lefebvre, P., Silva, U.~L., and Prins, E. (2001).
\newblock Road paving, fire regime feedbacks, and the future of amazon forests.
\newblock {\em Forest ecology and management}, 154(3):395--407.

\bibitem[Nepstad et~al., 1999]{nepstad1999large}
Nepstad, D.~C., Verssimo, A., Alencar, A., Nobre, C., Lima, E., Lefebvre, P.,
  Schlesinger, P., Potter, C., Moutinho, P., Mendoza, E., et~al. (1999).
\newblock Large-scale impoverishment of amazonian forests by logging and fire.
\newblock {\em Nature}, 398(6727):505--508.

\bibitem[Oliveira and Marquis, 2002]{oliveira2002cerrados}
Oliveira, P.~S. and Marquis, R.~J. (2002).
\newblock {\em The Cerrados of Brazil: Ecology and Natural History of a
  Neotropical Savanna}.
\newblock Columbia University Press.

\bibitem[Pan et~al., 2011]{pan2011large}
Pan, Y., Birdsey, R.~A., Fang, J., Houghton, R., Kauppi, P.~E., Kurz, W.~A.,
  Phillips, O.~L., Shvidenko, A., Lewis, S.~L., Canadell, J.~G., et~al. (2011).
\newblock A large and persistent carbon sink in the world's forests.
\newblock {\em Science}, 333(6045):988--993.

\bibitem[Pellegrini et~al., 2016]{pellegrini2016shifts}
Pellegrini, A.~F., Franco, A.~C., and Hoffmann, W.~A. (2016).
\newblock Shifts in functional traits elevate risk of fire-driven tree dieback
  in tropical savanna and forest biomes.
\newblock {\em Global change biology}, 22(3):1235--1243.

\bibitem[Pueyo et~al., 2010]{pueyo2010testing}
Pueyo, S., De~Alencastro~Gra{\c{c}}a, P. M.~L., Barbosa, R.~I., Cots, R.,
  Cardona, E., and Fearnside, P.~M. (2010).
\newblock Testing for criticality in ecosystem dynamics: the case of amazonian
  rainforest and savanna fire.
\newblock {\em Ecology Letters}, 13(7):793--802.

\bibitem[Romme and Despain, 1989]{romme1989historical}
Romme, W.~H. and Despain, D.~G. (1989).
\newblock Historical perspective on the yellowstone fires of 1988.
\newblock {\em BioScience}, 39(10):695--699.

\bibitem[Rossatto et~al., 2009]{rossatto2009differences}
Rossatto, D.~R., Hoffmann, W.~A., and Franco, A.~C. (2009).
\newblock Differences in growth patterns between co-occurring forest and
  savanna trees affect the forest--savanna boundary.
\newblock {\em Functional Ecology}, 23(4):689--698.

\bibitem[Scheffer et~al., 2001]{scheffer2001catastrophic}
Scheffer, M., Carpenter, S., Foley, J.~A., Folke, C., and Walker, B. (2001).
\newblock Catastrophic shifts in ecosystems.
\newblock {\em Nature}, 413(6856):591--596.

\bibitem[Schertzer et~al., 2015]{schertzer2015implications}
Schertzer, E., Staver, A., and Levin, S. (2015).
\newblock Implications of the spatial dynamics of fire spread for the
  bistability of savanna and forest.
\newblock {\em Journal of Mathematical Biology}, 70(1-2):329--341.

\bibitem[Smith and Schindler, 2009]{Smith2009201}
Smith, V.~H. and Schindler, D.~W. (2009).
\newblock Eutrophication science: where do we go from here?
\newblock {\em Trends in Ecology \& Evolution}, 24(4):201 -- 207.

\bibitem[Staver et~al., 2011]{staver2011global}
Staver, A.~C., Archibald, S., and Levin, S.~A. (2011).
\newblock The global extent and determinants of savanna and forest as
  alternative biome states.
\newblock {\em Science}, 334(6053):230--232.

\bibitem[Tilman et~al., 1994]{tilman1994habitat}
Tilman, D., May, R.~M., Lehman, C.~L., and Nowak, M.~A. (1994).
\newblock Habitat destruction and the extinction debt.
\newblock {\em Nature}, 371:65--66.

\bibitem[Turner, 1996]{turner1996species}
Turner, I. (1996).
\newblock Species loss in fragments of tropical rain forest: a review of the
  evidence.
\newblock {\em Journal of applied Ecology}, pages 200--209.

\bibitem[Turner, 2005]{turner2005landscape}
Turner, M.~G. (2005).
\newblock Landscape ecology: what is the state of the science?
\newblock {\em Annu. Rev. Ecol. Evol. Syst.}, 36:319--344.

\bibitem[Turner, 2010]{turner2010disturbance}
Turner, M.~G. (2010).
\newblock Disturbance and landscape dynamics in a changing world.
\newblock {\em Ecology}, 91(10):2833--2849.

\bibitem[Uhl and Kauffman, 1990]{uhl1990deforestation}
Uhl, C. and Kauffman, J.~B. (1990).
\newblock Deforestation, fire susceptibility, and potential tree responses to
  fire in the eastern amazon.
\newblock {\em Ecology}, 71(2):437--449.

\bibitem[Yassemi et~al., 2008]{yassemi2008design}
Yassemi, S., Dragi{\'c}evi{\'c}, S., and Schmidt, M. (2008).
\newblock Design and implementation of an integrated gis-based cellular
  automata model to characterize forest fire behaviour.
\newblock {\em ecological modelling}, 210(1):71--84.

\end{thebibliography}

\end{document}